\documentstyle[twoside,fleqn,epsfig]{article}

\def\be{\begin{equation}}
\def\ee{\end{equation}}
\def\bea{\begin{eqnarray}}
\def\eea{\end{eqnarray}}
\newcommand{\lsim}{\mathrel{\mathop{\kern 0pt \rlap
  {\raise.2ex\hbox{$<$}}}
  \lower.9ex\hbox{\kern-.190em $\sim$}}}
\newcommand{\gsim}{\mathrel{\mathop{\kern 0pt \rlap
  {\raise.2ex\hbox{$>$}}}
  \lower.9ex\hbox{\kern-.190em $\sim$}}}

\newcommand{\AmS}{{\protect\the\textfont2
  A\kern-.1667em\lower.5ex\hbox{M}\kern-.125emS}}
\normalsize
\begin{document}

\baselineskip=0.65cm

\begin{center}
\Large
{\bf New results from DAMA/LIBRA}
\rm
\end{center}

\large

\begin{center}
R.\,Bernabei $^{a,b}$,~P.\,Belli $^{b}$, ~F.\,Cappella $^{c,d}$, R.\,Cerulli $^{e}$, C.J.\,Dai $^{f}$,
\vspace{1mm}

A. d'Angelo $^{c,d}$, ~H.L.\,He $^{f}$, ~A.\,Incicchitti $^{d}$, ~H.H.\,Kuang $^{f}$,
\vspace{1mm}

~X.H.\,Ma $^{f}$, ~F.\,Montecchia $^{a,b}$, ~F.\,Nozzoli $^{a,b}$,
\vspace{1mm}

~D.\,Prosperi $^{c,d}$, ~X.D.\,Sheng $^{f}$, ~R.G.\,Wang $^{f}$, ~Z.P.\,Ye $^{f,g}$
\vspace{1mm}

\normalsize
\vspace{0.4cm}

$^{a}${\it Dip. di Fisica, Universit\`a di Roma ``Tor Vergata'', I-00133  Rome, Italy}
\vspace{1mm}

$^{b}${\it INFN, sez. Roma ``Tor Vergata'', I-00133 Rome, Italy}
\vspace{1mm}

$^{c}${\it Dip. di Fisica, Universit\`a di Roma ``La Sapienza'', I-00185 Rome, Italy}
\vspace{1mm}

$^{d}${\it INFN, sez. Roma, I-00185 Rome, Italy}
\vspace{1mm}

$^{e}${\it Laboratori Nazionali del Gran Sasso, I.N.F.N., Assergi, Italy}
\vspace{1mm}

$^{f}${\it IHEP, Chinese Academy, P.O. Box 918/3, Beijing 100039, China}
\vspace{1mm}

$^{g}${\it University of Jing Gangshan, Jiangxi, China}
\vspace{1mm}

\end{center}

\vspace{0.2cm}
\normalsize

\begin{abstract}

DAMA/LIBRA is running at the Gran Sasso National Laboratory 
of the I.N.F.N.. Here the results obtained with a further exposure 
of 0.34 ton $\times$ yr are presented. They refer to two further annual 
cycles collected one before and one after the first DAMA/LIBRA upgrade occurred 
on September/October 2008.  The cumulative 
exposure with those previously released by the former DAMA/NaI and by DAMA/LIBRA is
now 1.17 ton $\times$ yr, corresponding to 13 annual cycles. 
The data further confirm the model independent evidence 
of the presence of Dark Matter (DM) particles in the galactic halo on the basis 
of the DM annual modulation signature (8.9 $\sigma$ C.L. for the cumulative exposure). 
In particular, with the cumulative exposure
the modulation amplitude of the {\it single-hit} events in the (2 -- 6) keV energy 
interval measured in NaI(Tl) target is ($0.0116 \pm 0.0013$) 
cpd/kg/keV; the measured phase is ($146 \pm 7$) days and the measured period is
($0.999 \pm 0.002$) yr, values well
in agreement with those expected for the DM particles.
\end{abstract}

\vspace{5.0mm}


\vspace{2.0mm}


\section{Introduction}

The former DAMA/NaI \cite{prop,allDM,Nim98,Sist,RNC,ijmd,ijma,epj06,ijma07,chan,wimpele,ldm,allRare,IDM96} 
and the present DAMA/LIBRA \cite{modlibra,perflibra,papep} 
experiments at the Gran Sasso National Laboratory have the main aim to investigate the 
presence of Dark Matter particles in the galactic halo by exploiting the model independent 
Dark Matter annual modulation signature originally suggested in the mid 80's 
in ref. \cite{Freese}. 
In fact, as a consequence of its annual revolution around the Sun, which is moving in the Galaxy
travelling with respect to the Local Standard of Rest towards 
the star Vega near the constellation of Hercules, 
the Earth should be crossed by a larger flux of Dark Matter particles around $\sim$ 2 June
(when the Earth orbital velocity is summed to the one of the solar system with respect 
to the Galaxy) and by a smaller one around $\sim$ 2 December (when the two velocities are subtracted).
Thus, this signature has a different origin and peculiarities than the seasons on the Earth
and than effects correlated with seasons (consider the expected value of the 
phase as well as the other requirements listed below). 
This annual modulation signature is very distinctive since the effect induced by DM
particles must simultaneously satisfy all the following requirements:
the rate must contain a component
modulated according to a cosine function (1) with one year period (2)
and a phase that peaks roughly around $\simeq$ 2$^{nd}$ June (3);
this modulation must only be found
in a well-defined low energy range, where DM particle induced events
can be present (4); it must apply only to those events in
which just one detector of many actually ``fires'' ({\it single-hit} events), since
the DM particle multi-interaction probability is negligible (5); the modulation
amplitude in the region of maximal sensitivity must be $\lsim$7$\%$
for usually adopted halo distributions (6), but it can
be larger in case of some possible scenarios such as e.g. those in refs. \cite{Wei01,Fre04}. 
This offers an efficient DM model independent signature, able to test a large interval of
cross sections and of halo densities; moreover, the use of highly 
radiopure NaI(Tl) scintillators as target-detectors assures sensitivity to wide 
ranges of DM candidates, of interaction types and of astrophysical scenarios.
 
It is worth noting that only systematic effects or side reactions 
able to simultaneously fulfil all the 6 requirements given above (and no one has ever been
suggested) and to account for the whole observed modulation amplitude might mimic 
this DM signature.

\vspace{0.5cm}

The DAMA/LIBRA set-up, whose description, radiopurity and main
features are discussed in details in ref. \cite{perflibra} has firstly been upgraded 
in September/October 2008: 
i) one detector has been recovered by replacing a broken PMT (see ref. \cite{modlibra});
ii) a new optimization of some PMTs and HVs has been performed; iii)
all the transient digitizers recording the shape of the pulse have been replaced 
with new ones, the U1063A Acqiris 8-bit 1GS/s DC270 High-Speed cPCI Digitizers; 
iv) a new DAQ with optical read-out has been installed. 
Also during this upgrade the operations involving the handling of the
sensitive
part of the setup and the shield have been performed in HP Nitrogen atmosphere.  
The upgrade has allowed to enlarge the sensitive mass and to improve general features.
Here we just remind that    
the sensitive part of this set-up is made of 25 highly radiopure NaI(Tl) crystal scintillators
(5-rows by 5-columns matrix) having 9.70 kg mass each one.
In each detector two 10 cm long special quartz light guides act also as
optical windows on the two end faces of the crystal and are coupled to two low background
photomultipliers working in coincidence at single photoelectron level. The detectors are 
housed in a sealed low-radioactive
copper box installed in the center of a low-radioactive Cu/Pb/Cd-foils/polyethylene/paraffin shield;
moreover, about 1 m concrete (made from the Gran Sasso rock material) almost fully surrounds (mostly
outside the barrack) this passive shield, acting as a further neutron moderator.
A threefold-levels sealing system excludes the detectors from the environmental air of the 
underground laboratory \cite{perflibra}.  
A hardware/software system to monitor the running conditions is operative and self-controlled
computer processes automatically control several parameters and manage alarms.
Moreover: i) the light response ranges 
from 5.5 to 7.5 photoelectrons/keV, depending on the detector; ii) the hardware threshold 
of each PMT is at single photoelectron (each detector is equipped with two low background 
photomultipliers working in coincidence); 
iii) energy calibration with X-rays/$\gamma$ sources are regularly carried out down to few keV; 
iv) the software energy threshold of the experiment is 2 keV;
v) both {\it single-hit} events (where just one of the detectors fires) and 
{\it multiple-hit} 
events (where more than one detector fires) are acquired;
v) the data are collected up to the MeV region despite the optimization is performed for 
the lower one. 
For the radiopurity, the procedures and further
details see ref. \cite{modlibra,perflibra}.

The data of the former DAMA/NaI (0.29 ton $\times$ yr) and 
those of the first 4 annual cycles of DAMA/LIBRA 
(total exposure 0.53 ton$\times$yr) 
have already given positive model independent evidence for the presence of DM particles in 
the galactic halo with high confidence level on the basis of the DM annual modulation signature
\cite{modlibra}.

In this paper the model independent results with other two annual cycles DAMA/LIBRA-5,6
are presented. As mentioned, the data of the first cycle have been collected  in the same 
conditions as 
DAMA/LIBRA-1,2,3,4 \cite{modlibra,perflibra}, while the data of DAMA/LIBRA-6
have been taken after the above mentioned 2008 upgrade.

\section{The results}

The updated exposures of the DAMA/LIBRA annual cycles and the cumulative one 
with the former DAMA/NaI are given in Table \ref{tb:years}.

\begin{table}[ht]
\caption{Exposures of the DAMA/LIBRA-5,6 annual cycles. Here $\alpha=\langle 
cos^2\omega (t-t_0) \rangle$
is the mean value of the squared cosine and $\beta=\langle cos \omega (t-t_0) \rangle$
is the mean value of the cosine (the averages are taken over the live time of the data taking
and $t_0=152.5$ day, i.e. June 2$^{nd}$);
thus, $(\alpha - \beta^2)$ indicates the variance of the cosine
(i.e. it is 0.5 for a full year of data taking).
The information on the previously published DAMA/LIBRA-1,2,3,4,
are recalled as well as the cumulative exposure, when 
including the former DAMA/NaI.}
\vspace{-0.5cm}
\begin{center}
\resizebox{\textwidth}{!}{
\begin{tabular}{|c|c|c|c|c|}
\hline
 & Period  & mass & Exposure  & $(\alpha - \beta^2)$ \\
 &         &  (kg)& (kg$\times$day) &  \\
\hline
 DAMA/LIBRA-1,2,3,4 & Sept. 9, 2003 - July 17, 2007 & 232.8 & 192824 & 0.537 \\
   &        &       & & \\
 DAMA/LIBRA-5 & July 17, 2007 - Aug. 29, 2008 & 232.8 & 66105  & 0.468 \\
   &        &       & & \\
 DAMA/LIBRA-6 & Nov. 12,2008 - Sept. 1, 2009 & 242.5 &58768 & 0.519 \\
   &        &       & & \\
\hline
 DAMA/LIBRA-1 to -6 & Sept. 9, 2003 - Sept. 1, 2009 & &317697$\simeq$ 
 0.87 ton$\times$yr & 0.519 \\
\hline
\multicolumn{3}{|l}{DAMA/NaI + DAMA/LIBRA-1 to 6:} & \multicolumn{2}{c|}{1.17 ton$\times$yr}  \\
\hline
\hline
\end{tabular}}
\end{center}
\label{tb:years}
\end{table}

The only data treatment, which is performed on the raw data, is to remove noise pulses
(mainly PMT noise, Cherenkov light in the light guides and in the PMT windows, and afterglows)
near the energy threshold in the {\it single-hit} events;
for a description of the used procedure and details see ref. \cite{perflibra}.

In the DAMA/LIBRA-1,2,3,4,5,6 annual cycles about $7.2 \times 10^7$ events have 
also been collected for energy calibrations
and about $3 \times 10^6$ events/keV for the evaluation of the acceptance windows 
efficiency for noise rejection near energy threshold.
The periodical calibrations and, in particular, those related with the acceptance windows efficiency
mainly affect the duty cycle of the experiment.
From Table \ref{tb:years} one can observe a significant improvement
in the duty cycle of the sixth annual cycles with respect to the previous ones;
this is mainly due to the new transient Digitizers and DAQ installed at fall 2008. 

\begin{figure}[!ht]
\begin{center}
\includegraphics[width=0.9\textwidth] {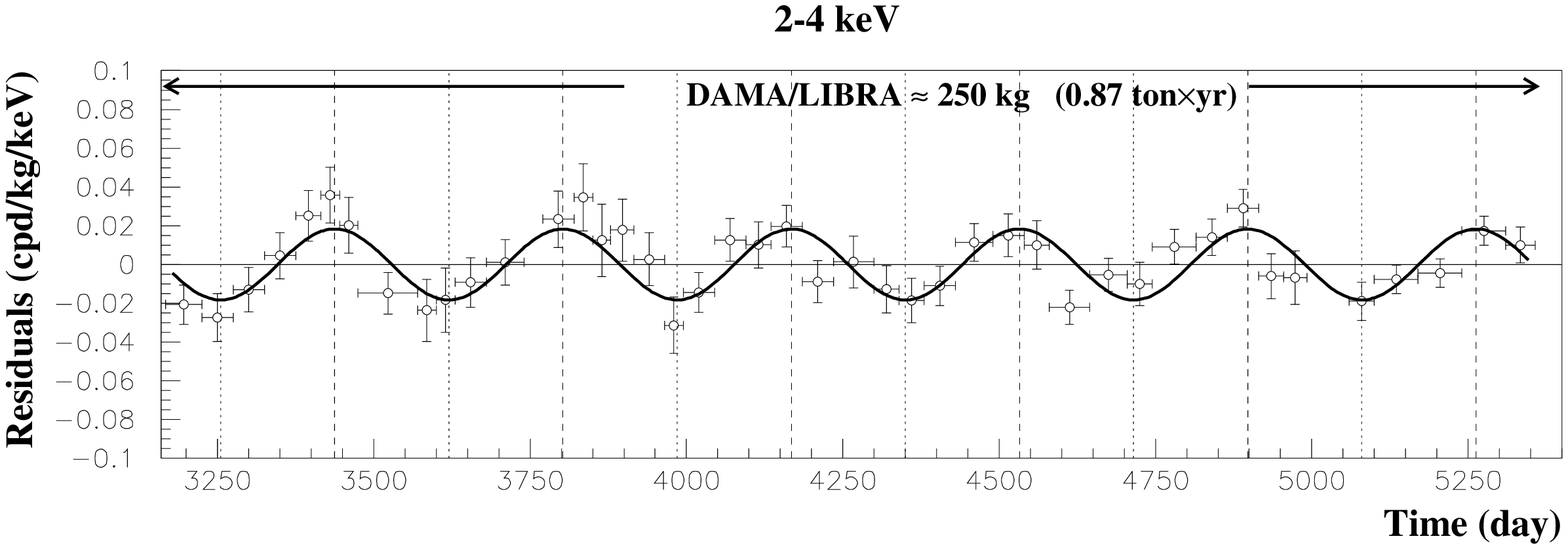}
\includegraphics[width=0.9\textwidth] {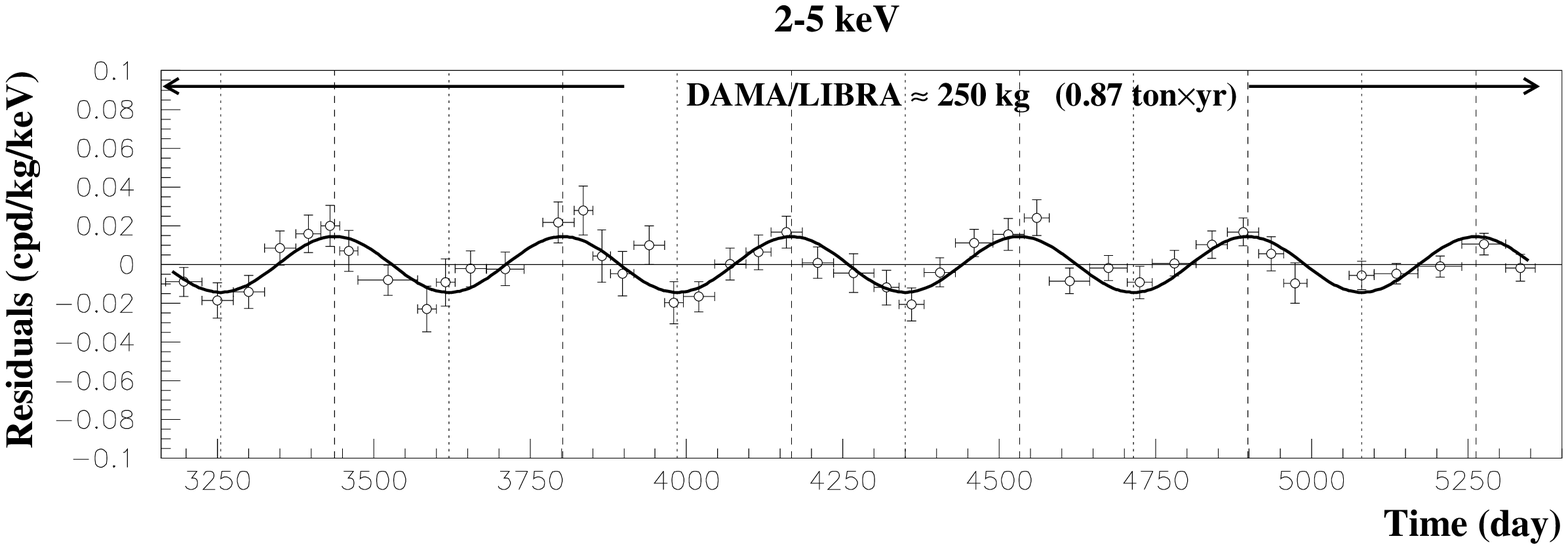} 
\includegraphics[width=0.9\textwidth] {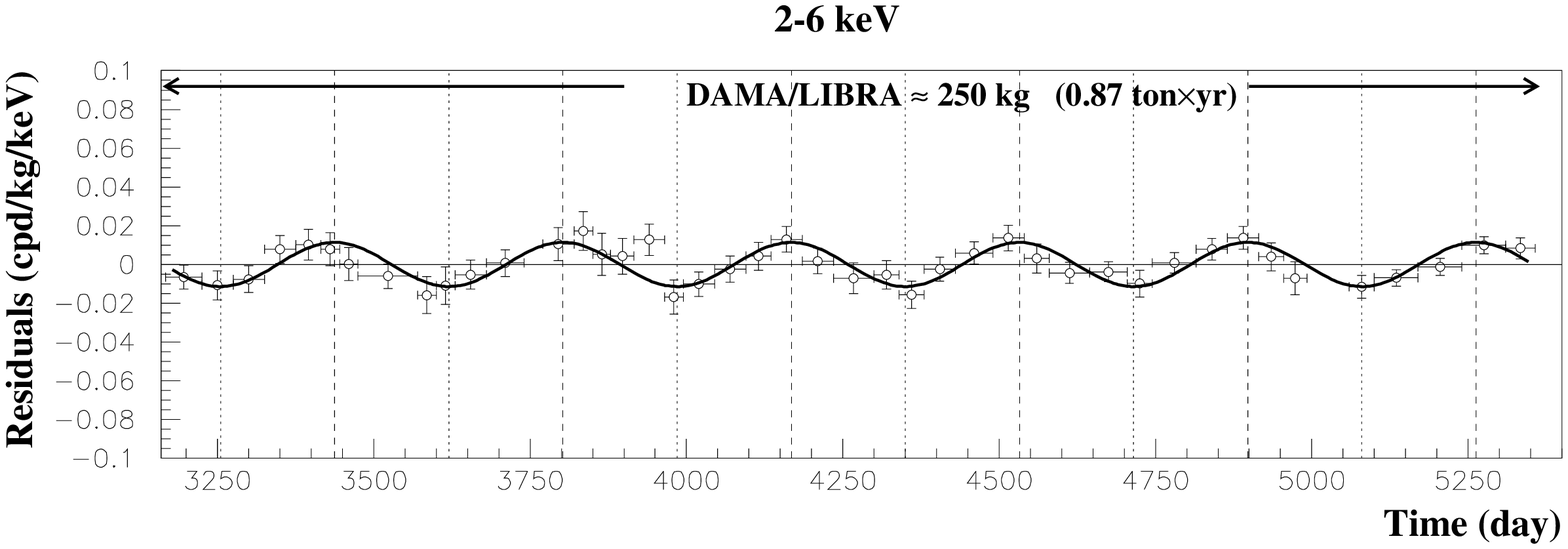}
\end{center}
\vspace{-0.4cm}
\caption{Experimental model-independent residual rate of the {\it single-hit} scintillation events, 
measured by DAMA/LIBRA,1,2,3,4,5,6 in the (2 -- 4), (2 -- 5) and (2 -- 6) keV energy intervals 
as a function of the time. The zero of the time scale is January 1$^{st}$ 
of the first year of data taking of the former DAMA/NaI experiment \cite{modlibra}.
The experimental points present the errors as vertical bars and the associated time bin width as horizontal bars. 
The superimposed curves are the cosinusoidal functions behaviors $A \cos \omega(t-t_0)$  
with a period $T = \frac{2\pi}{\omega} =  1$ yr, with a phase $t_0 = 152.5$ day (June 2$^{nd}$) and with
modulation amplitudes, $A$, equal to the central values obtained by best fit over the whole data 
including also the exposure previously collected by the former DAMA/NaI experiment: cumulative exposure 
is 1.17 ton $\times$ yr (see also ref. \cite{modlibra} and refs. therein). The dashed vertical lines 
correspond to the maximum expected for the DM signal (June 2$^{nd}$), while 
the dotted vertical lines correspond to the minimum. See text.}
\label{fg:res}
\vspace{-0.2cm}
\end{figure}

Several analyses on the model-independent investigation of the DM annual 
modulation signature have been performed as previously done in ref. \cite{modlibra} 
and refs. therein. In particular, Fig. \ref{fg:res} shows the time behaviour of the experimental 
residual rates for {\it single-hit} 
events in the (2--4), (2--5) and (2--6) keV energy intervals. These residual rates are 
calculated from the measured rate of the {\it single-hit} events (already corrected 
for the overall efficiency and for the acquisition dead time)
after subtracting the constant part: $<r_{ijk}-flat_{jk}>_{jk}$.
Here $r_{ijk}$ is the rate in the considered $i$-th time interval for the $j$-th detector in the 
$k$-th energy bin, while $flat_{jk}$ is the rate of the $j$-th detector in the $k$-th energy bin 
averaged over the cycles. 
The average is made on all the detectors ($j$ index) and on all the energy bins ($k$ index) which 
constitute the considered energy interval. The weighted mean of the residuals must 
obviously be zero over one cycle. 
 
For clarity in Fig. \ref{fg:res} only the DAMA/LIBRA data collected over 
six annual cycles (0.87 ton $\times$ yr) are shown; the DAMA/NaI data (0.29 ton $\times$ 
yr) and comparison with DAMA/LIBRA are available in ref. \cite{modlibra}. 

The hypothesis of absence of modulation in the data can be discarded (see Table \ref{tb:mod_0}). 

\begin{table}[ht]
\caption{Test of absence of modulation in the data of the
DAMA/LIBRA-1,2,3,4,5,6 and without/with 
also the data of the former DAMA/NaI. As it can be seen, a null modulation amplitude 
is discarded by the data.}
\begin{center}
\begin{tabular}{|c|c|c|}
\hline
 Energy interval & DAMA/LIBRA & DAMA/NaI \& DAMA/LIBRA  \\
 (keV) & (6 annual cycles) & (7+6 annual cycles) \\
\hline
 2-4 & $\chi^2$/d.o.f. = 90.0/43 & $\chi^2$/d.o.f. = 147.4/80 \\
  & $\rightarrow$ P = 3.6 $\times$ 10$^{-5}$&$\rightarrow$ P = 6.8 $\times$ 10$^{-6}$ \\
\hline
 2-5 & $\chi^2$/d.o.f. = 82.1/43               & $\chi^2$/d.o.f. = 135.2/80 \\
     & $\rightarrow$ P = 3.1 $\times 10^{-4}$  & $\rightarrow$ P = 1.1 $\times$ 10$^{-4}$ \\
\hline
 2-6 & $\chi^2$/d.o.f. = 68.9/43 & $\chi^2$/d.o.f. = 139.5/80 \\
  & $\rightarrow$ P = 7.4 $\times$ 10$^{-3}$&$\rightarrow$ P = 4.3 $\times$ 10$^{-5}$ \\
\hline
\hline
\end{tabular}
\end{center}
\label{tb:mod_0}
\end{table}

The {\it single-hit} residual rate  of DAMA/LIBRA-1,2,3,4,5,6 of Fig. \ref{fg:res} can be 
fitted with the formula: $A \cos \omega(t-t_0)$ considering a
period $T = \frac{2\pi}{\omega} =  1$ yr and a phase $t_0 = 152.5$ day (June 2$^{nd}$), as 
expected by the DM annual modulation signature; this can be repeated for the total 
available exposure 1.17 ton $\times$ yr including the former 
DAMA/NaI data (see \cite{modlibra} and refs. therein).
The results are shown in Table \ref{tb:ampff}.

\begin{table}[ht]
\caption{Modulation amplitude, A, obtained by fitting the {\it single-hit} residual rate  
of the six DAMA/LIBRA annual cycles (Fig. \ref{fg:res}), and including also 
the former DAMA/NaI data given elsewhere (see \cite{modlibra} and refs. therein)
for a total cumulative exposure of 1.17 ton $\times$ yr. It has been obtained by fitting
the data with the formula: 
$A \cos \omega(t-t_0)$ with $T = \frac{2\pi}{\omega} =  1$ yr and $t_0 = 152.5$ day (June 
2$^{nd}$), as
expected for a signal by the DM annual modulation signature. The corresponding $\chi^2$
value for each fit and the confidence level are also reported}
\begin{center}
\begin{tabular}{|c|c|c|}
\hline
 Energy interval & DAMA/LIBRA & DAMA/NaI \& DAMA/LIBRA  \\
 (keV) & (cpd/kg/keV) & (cpd/kg/keV) \\
\hline
 2-4 & A=(0.0170$\pm$0.0024) & A=(0.0183$\pm$0.0022) \\
  & $\chi^2$/d.o.f. = 41.0/42 & $\chi^2$/d.o.f. = 75.7/79 \\
  &   & $\rightarrow$ 8.3 $\sigma$ C.L. \\
\hline
 2-5 & A=(0.0129$\pm$0.0018) & A=(0.0144$\pm$0.0016) \\
  & $\chi^2$/d.o.f. = 30.7/42 & $\chi^2$/d.o.f. = 56.6/79 \\
  &   & $\rightarrow$ 9.0 $\sigma$ C.L. \\
\hline 
2-6 & A=(0.0097$\pm$0.0015) & A=(0.0114$\pm$0.0013) \\
  & $\chi^2$/d.o.f. = 24.1/42 & $\chi^2$/d.o.f. = 64.7/79 \\
  &   & $\rightarrow$ 8.8 $\sigma$ C.L. \\
\hline
\hline
\end{tabular}
\end{center}
\label{tb:ampff}
\end{table}

The compatibility among the 13 annual cycles has been investigated. In particular,
the modulation 
amplitudes measured in each annual cycle of the whole 1.17 ton $\times$ yr exposure 
have been analysed as in ref. \cite{modlibra}. Indeed these modulation amplitudes are normally 
distributed 
around their best fit value as pointed out by the $\chi^2$ test
($\chi^2 = 9.3$, 12.2 and 10.1 over 12 {\it d.o.f.} for the three energy 
intervals, respectively) and the {\it run test} (lower tail probabilities
of 57\%, 47\% and 35\% for the three energy intervals, respectively).
Moreover, the 
DAMA/LIBRA-5 and DAMA/LIBRA-6 (2--6) keV modulation amplitudes are $(0.0086\pm0.0032)$ cpd/kg/keV
and $(0.0101\pm0.0031)$ cpd/kg/keV, respectively, in agreement with that of 
DAMA/LIBRA-1,2,3,4: $(0.0110\pm0.0019)$ cpd/kg/keV; we also recall that 
the statistical compatibility between the DAMA/NaI and
DAMA/LIBRA-1,2,3,4 modulation amplitudes has been verified \cite{modlibra}.
Thus, also when adding DAMA/LIBRA-5,6, 
the cumulative result from DAMA/NaI and DAMA/LIBRA can be adopted.

Table \ref{tb:ampfv} shows the results obtained for the cumulative 
1.17 ton $\times$ yr exposure when the period and phase parameters are 
kept free in the fitting procedure described above.

\begin{table}[ht]
\caption{Modulation amplitude ($A$), period ($T = \frac{2\pi}{\omega}$)
and phase ($t_0$), obtained by fitting, with the formula: 
$A \cos \omega(t-t_0)$, the {\it single-hit} 
residual rate of the cumulative 1.17 ton $\times$ yr exposure.
The results are well compatible with expectations for a signal in the 
DM annual modulation signature.}
\begin{center}
\begin{tabular}{|c|c|c|c|c|}
\hline
 Energy interval & $A$ (cpd/kg/keV)& $T = \frac{2\pi}{\omega}$ (yr) &  $t_0$ (days) & C. L. \\
\hline
 2-4 & (0.0194$\pm$0.0022) & (0.996$\pm$0.002) & 136$\pm$7 & 8.8$\sigma$\\
 2-5 & (0.0149$\pm$0.0016) & (0.997$\pm$0.002) & 142$\pm$7 & 9.3$\sigma$\\
 2-6 & (0.0116$\pm$0.0013) & (0.999$\pm$0.002) & 146$\pm$7 & 8.9$\sigma$\\
\hline
\hline
\end{tabular}
\end{center}
\label{tb:ampfv}
\end{table}

\begin{figure}[!tbh]
\centering
\vspace{-0.4cm}
\includegraphics[width=6.cm] {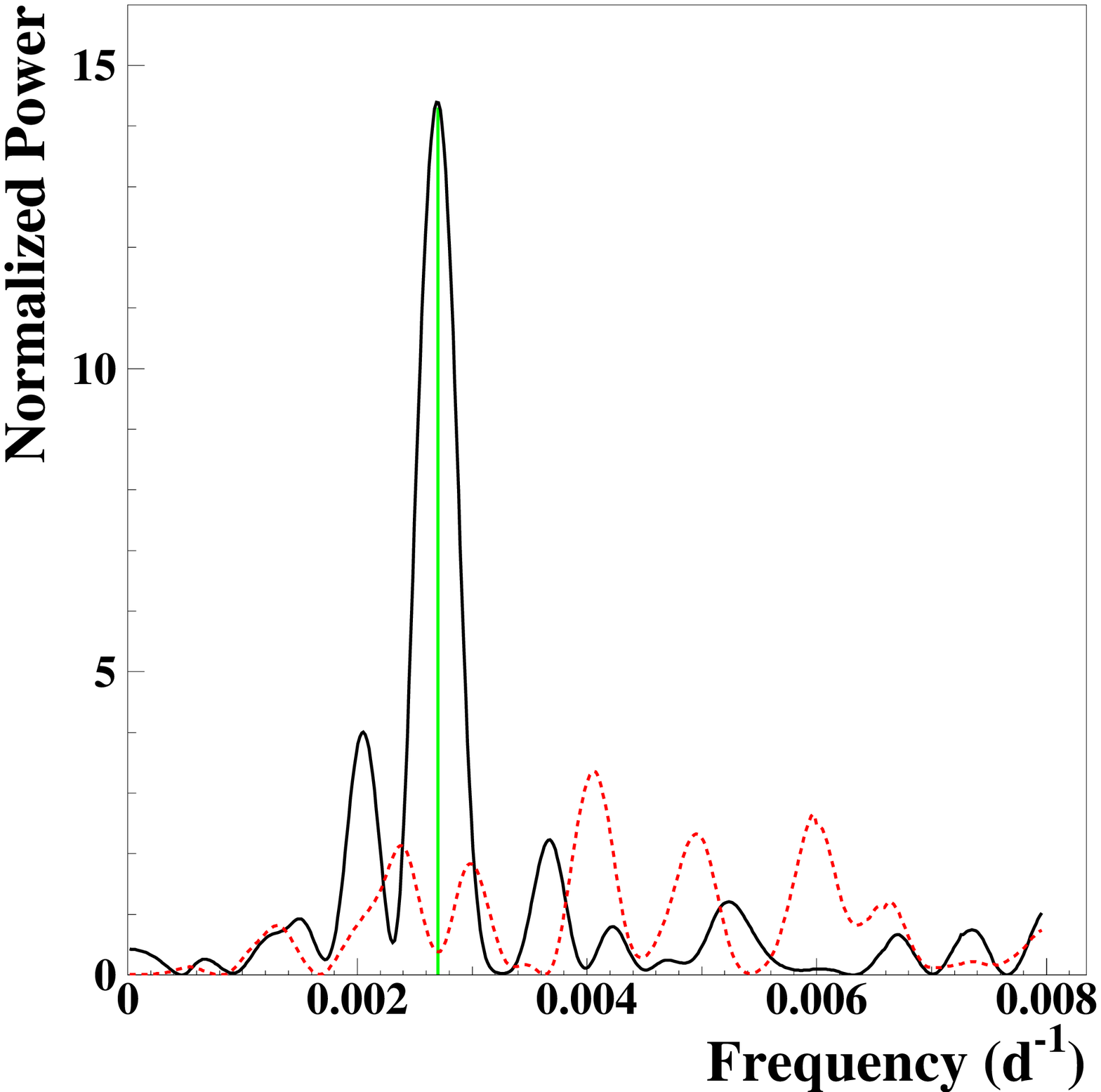}
\includegraphics[width=6.cm] {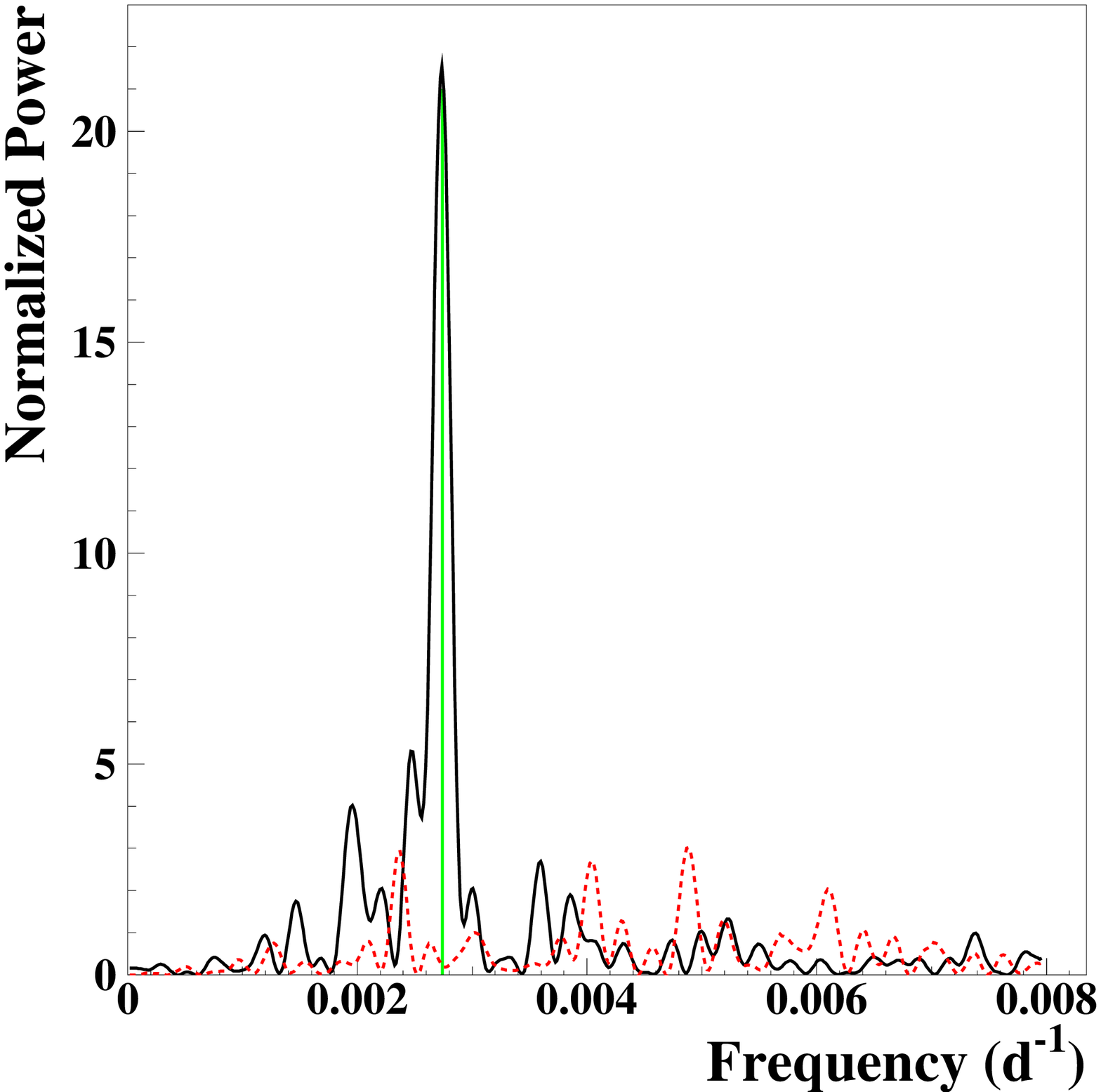}
\vspace{-0.3cm}
\caption{Power spectrum of the measured {\it single-hit} residuals in the (2--6) keV (solid lines) 
and (6--14) keV (dotted lines) energy intervals calculated according to ref. \cite{Lomb}, including also the
treatment of the experimental errors and of the time binning. The data refer to: a) DAMA/LIBRA-1,2,3,4,5,6
(exposure of 0.87 ton $\times$ yr); b) the cumulative 1.17 ton $\times$ yr
exposure (DAMA/NaI and DAMA/LIBRA-1,2,3,4,5,6).
As it can be seen, the principal mode present in the (2--6) keV energy interval corresponds to a frequency
of $2.697 \times 10^{-3}$ d$^{-1}$ and $2.735 \times 10^{-3}$ d$^{-1}$ 
(vertical lines), respectively in the a) and b) case. They correspond to a period 
of $\simeq$ 1 year. A similar peak is not present in the (6--14) keV energy interval just above.}
\label{fg:pwr}
\normalsize
\end{figure}

The DAMA/LIBRA {\it single-hit} residuals of Fig.\ref{fg:res} and those of DAMA/NaI 
(see 
e.g. \cite{modlibra}) have also been 
investigated by a Fourier analysis, obtaining a clear peak corresponding to a period of 1 year 
(see Fig. \ref{fg:pwr}); the same analysis in other energy region shows instead only aliasing peaks.

\vspace{0.3cm}

\begin{figure}[!ht]
\vspace{-0.5cm}
\begin{center}
\includegraphics[width=4.cm] {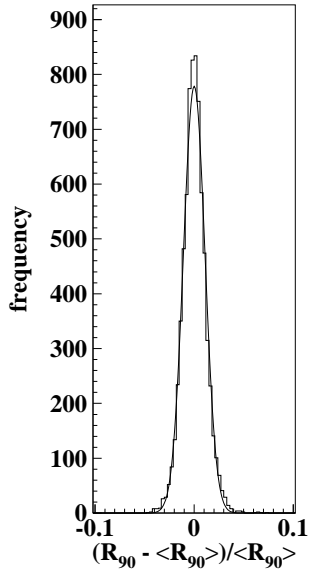}
\end{center}
\vspace{-0.5cm}
\caption{Distribution of the percentage variations
of R$_{90}$ with respect to the mean values for all the detectors in the DAMA/LIBRA-5,6 
annual cycles (histogram); the superimposed curve is a gaussian fit.}
\label{fig_r90}
\vspace{-0.2cm}
\end{figure}

\begin{figure}[!bth]
\vspace{-0.2cm}
\centering
a) \includegraphics[width=0.45\textwidth] {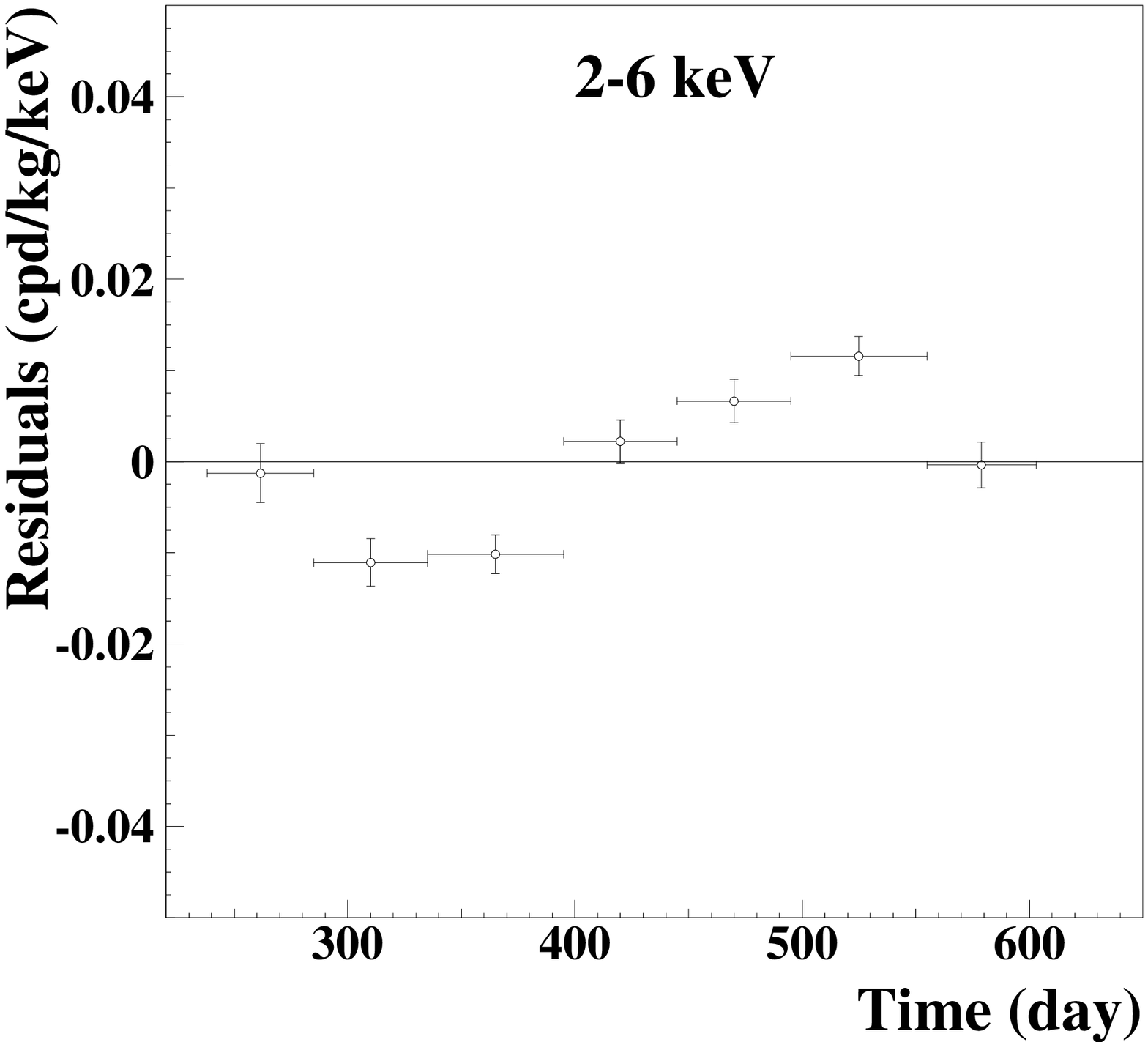}
b) \includegraphics[width=0.45\textwidth] {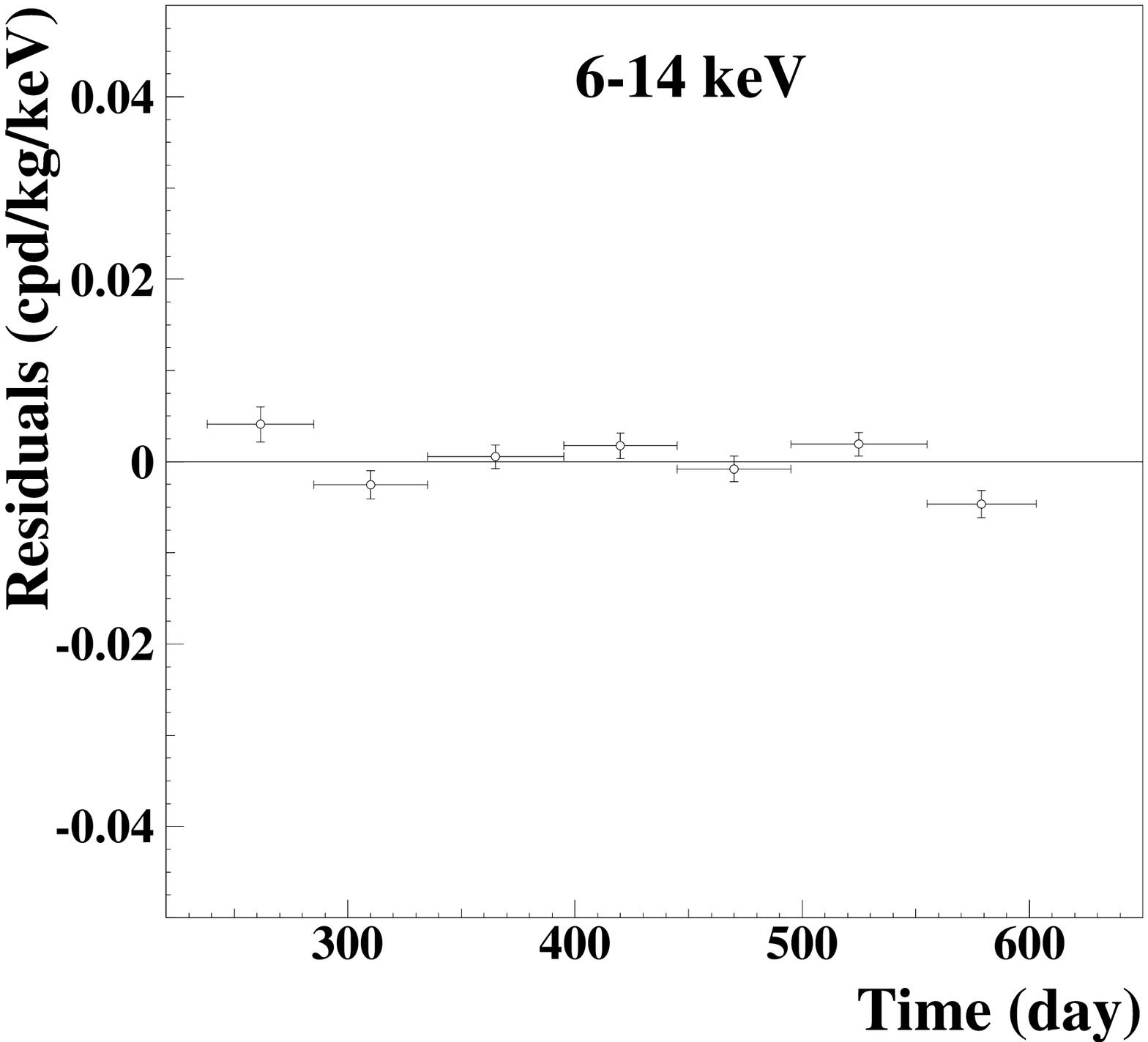}
\vspace{-0.2cm}
\caption{Experimental residuals in the (2 -- 6) keV region and those in the
(6 -- 14) keV energy region just above
for the cumulative 1.17 ton $\times$ yr,  considered as collected in a
single annual cycle. The experimental points present the errors as vertical bars and the associated
time bin width as horizontal bars. The initial time of the figure is taken at August 7$^{th}$.
The clear modulation satisfying all the peculiarities of the DM annual modulation signature is present 
in the lowest energy interval,
while it is absent just above; in fact, in the latter case 
the best fitted modulation amplitude is:
(0.00007 $\pm$ 0.00077) cpd/kg/keV.}
\label{fg:res1}
\vspace{-0.4cm}
\end{figure}

The measured energy distribution has been investigated in
other energy regions not of interest for Dark Matter,
also verifying the absence of any significant background modulation 
\footnote{,
In fact, the background in the lowest energy region is
essentially due to ``Compton'' electrons, X-rays and/or Auger
electrons, muon induced events, etc., which are strictly correlated
with the events in the higher energy part of the spectrum.
Thus, if a modulation detected
in the lowest energy region would be due to
a modulation of the background (rather than to a signal),
an equal or larger 
modulation in the higher energy regions should be present.}.
Following the procedures described 
in ref. \cite{modlibra} and ref. therein, the measured rate
integrated above 90 keV, R$_{90}$, as a function of the time has been analysed.
In particular, also for these two latter annual cycles the distribution of the percentage variations
of R$_{90}$ with respect to the mean values for all the detectors
has been considered; it shows a cumulative gaussian behaviour
with $\sigma$ $\simeq$ 1\%, well accounted by the statistical
spread expected from the used sampling time (see Fig. \ref{fig_r90}).
Moreover, fitting the time behaviour of R$_{90}$
with phase and period as for DM particles, a modulation amplitude compatible with zero
is also found in DAMA/LIBRA-5 and DAMA/LIBRA-6:
$(0.20 \pm 0.18)$ cpd/kg and $(-0.20 \pm 0.16)$ cpd/kg, respectively.
This also excludes the presence of any background
modulation in the whole energy spectrum at a level much
lower than the effect found in the lowest energy region for the {\it single-hit} events.
In fact, otherwise -- considering the R$_{90}$ mean values --
a modulation amplitude of order of tens
cpd/kg, that is $\simeq$ 100 $\sigma$ far away from the measured value, would be present.
Similar result is obtained when comparing 
the {\it single-hit} residuals in the (2--6) keV with those 
in other energy intervals; see as an example Fig. \ref{fg:res1}. 
It is worth noting that the obtained results already account for whatever 
kind of background and, in addition, that no background process able to mimic
the DM annual modulation signature (that is able to simultaneously satisfy 
all the peculiarities of the signature and to account for the measured modulation amplitude)
is available (see also discussions e.g. in \cite{modlibra,scineghe09}).

\begin{figure}[!ht]
\begin{center}
\includegraphics[width=0.9\textwidth] {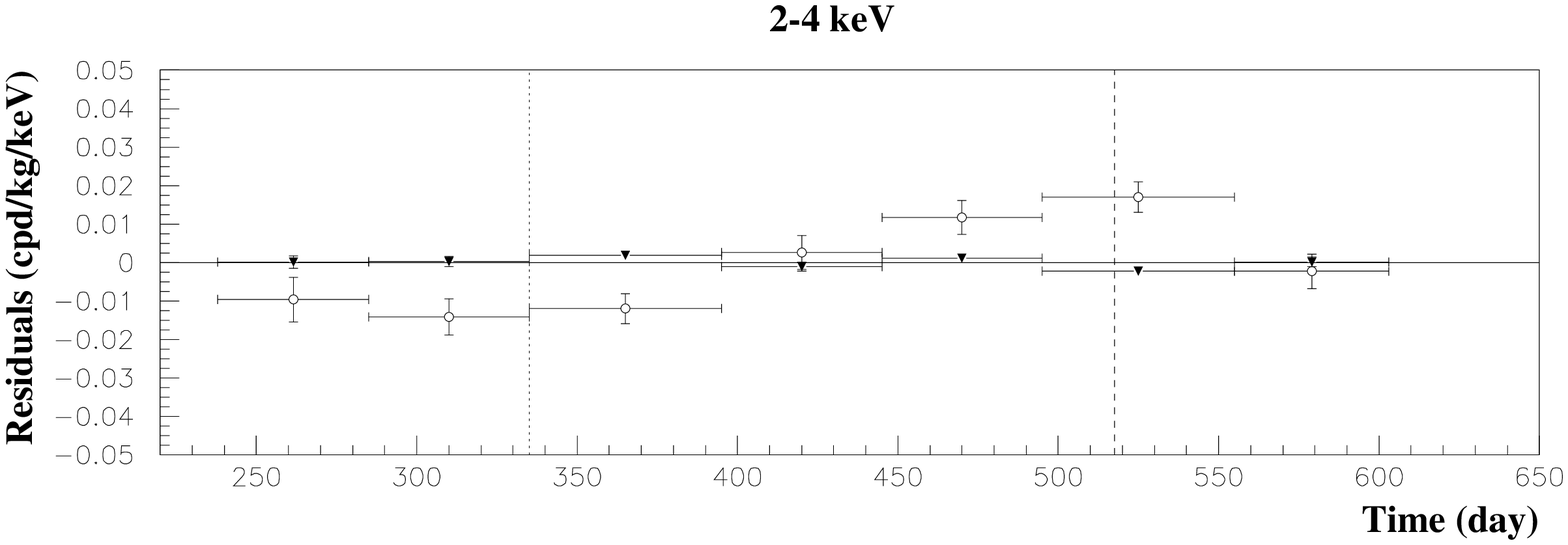}
\includegraphics[width=0.9\textwidth] {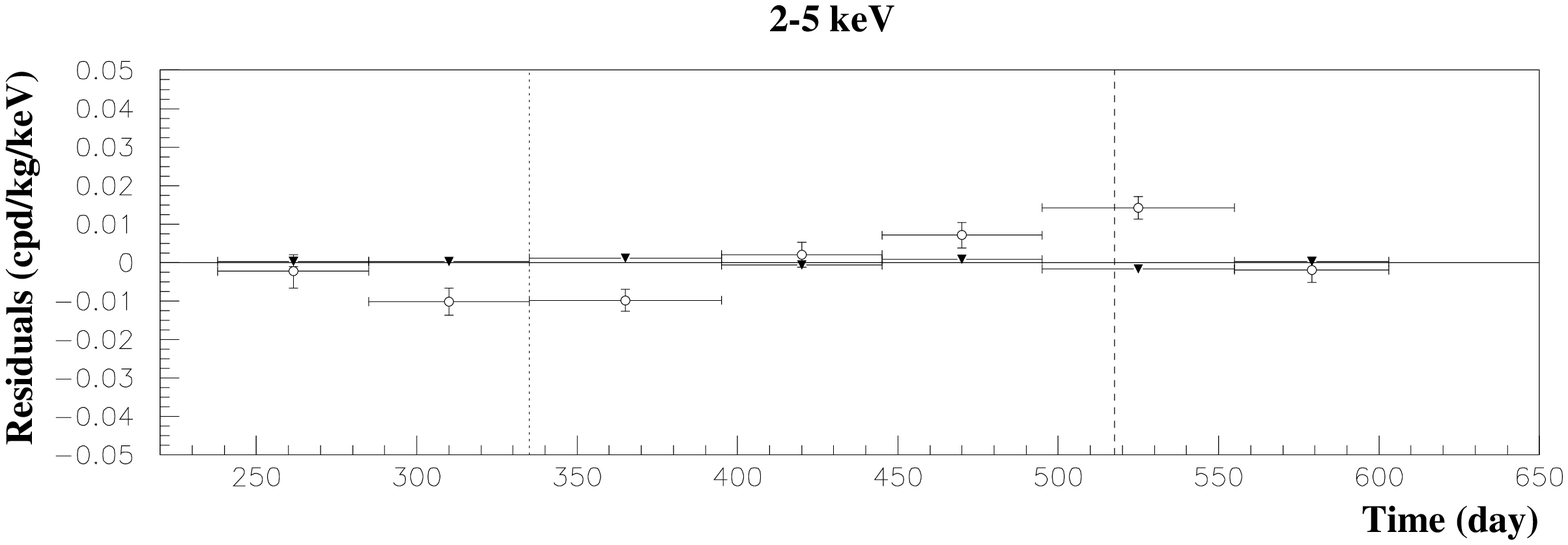}
\includegraphics[width=0.9\textwidth] {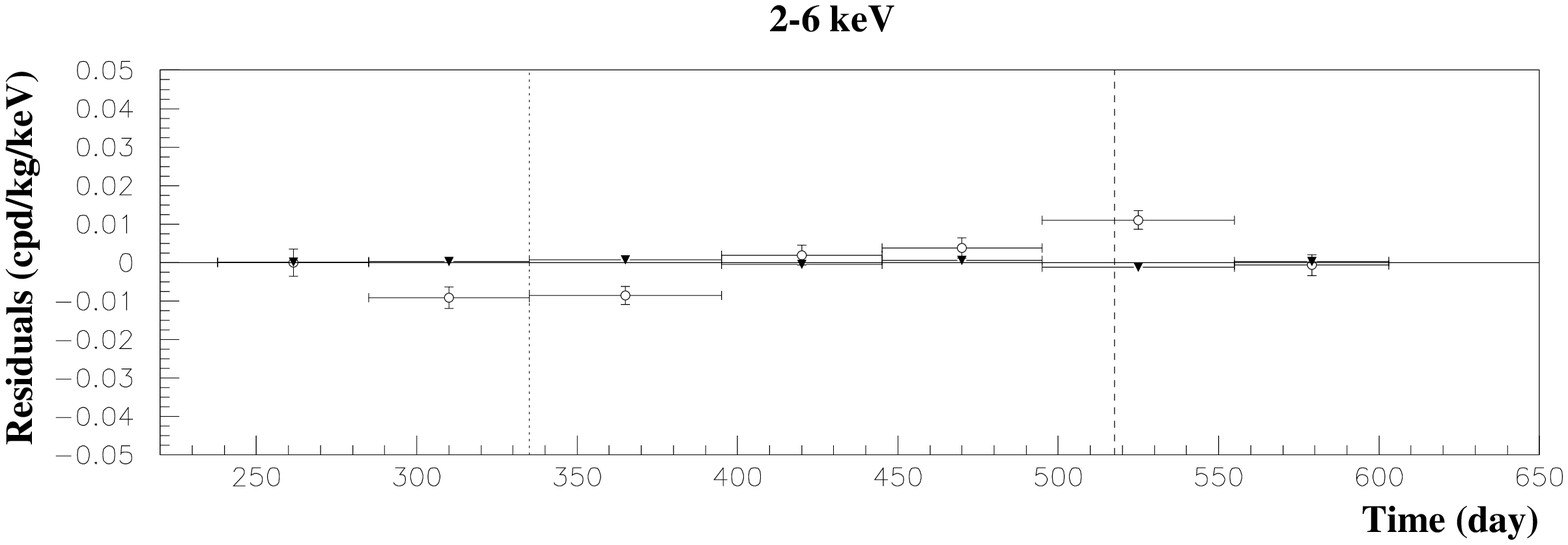}
\end{center}
\vspace{-0.5cm}
\caption{Experimental residual rates over the six DAMA/LIBRA annual cycles for {\it single-hit} events 
(open 
circles) (class of events to which DM events belong) and for {\it multiple-hit} events (filled triangles)
(class of events to which DM events do not belong).
They have been obtained by considering for each class of events the data as collected in a 
single annual cycle 
and by using in both cases the same identical hardware and the same identical software procedures.
The initial time of the figure is taken on August 7$^{th}$.
The experimental points present the errors as vertical bars and the associated time bin width as horizontal 
bars. See text and ref. \cite{modlibra}. 
Analogous results were obtained for the DAMA/NaI data 
\cite{ijmd}.}
\label{fig_mul}
\vspace{-0.4cm}
\end{figure}

A further relevant investigation has been performed by applying the same hardware and software 
procedures, used to acquire and to analyse the {\it single-hit} residual rate, to the 
{\it multiple-hit} one. 
In fact, since the probability that a DM particle interacts in more than one detector 
is negligible, a DM signal can be present just in the {\it single-hit} residual rate.
Thus, the comparison of the results of the {\it single-hit} events with those of the  {\it 
multiple-hit} ones corresponds practically to compare between them the cases of DM particles beam-on 
and beam-off.
This procedure also allows an additional test of the background behaviour in the same energy interval 
where the positive effect is observed. 
In particular, in Fig. \ref{fig_mul} the residual rates of the {\it single-hit} events measured over 
the six DAMA/LIBRA annual
cycles are reported, as collected in a single cycle, together with the residual rates 
of the {\it multiple-hit} events, in the considered energy intervals.
While, as already observed, a clear modulation, satisfying all the peculiarities of the DM
annual modulation signature, is present in 
the {\it single-hit} events,
the fitted modulation amplitudes for the {\it multiple-hit}
residual rate are well compatible with zero:
$(-0.0011\pm0.0007)$ cpd/kg/keV,
$(-0.0008\pm0.0005)$ cpd/kg/keV,
and $(-0.0006\pm0.0004)$ cpd/kg/keV
in the energy regions (2 -- 4), (2 -- 5) and (2 -- 6) keV, respectively.
Thus, again evidence of annual modulation with proper features as required by the DM annual 
modulation signature is present in the {\it single-hit} residuals (events class to which the
DM particle induced events belong), while it is absent in the {\it multiple-hit} residual 
rate (event class to which only background events belong).
Similar results were also obtained for the last two annual 
cycles of the
DAMA/NaI experiment \cite{ijmd}.
Since the same identical hardware and the same identical software procedures have been used to 
analyse the
two classes of events, the obtained result offers an additional strong support for the 
presence of a DM
particle component in the galactic halo.

As in ref. \cite{modlibra}, the annual modulation present at low energy can also 
be shown by depicting -- as a function of the energy --
the modulation amplitude, $S_{m,k}$, obtained
by maximum likelihood method over the data considering $T=$1 yr and $t_0=$ 152.5 day.
For such purpose the likelihood function of the {\it single-hit} experimental data
in the $k-$th energy bin is defined as: $ {\it\bf L_k}  = {\bf \Pi}_{ij} e^{-\mu_{ijk}}
{\mu_{ijk}^{N_{ijk}} \over N_{ijk}!}$,
where $N_{ijk}$ is the number of events collected in the
$i$-th time interval (hereafter 1 day), by the $j$-th detector and in the
$k$-th energy bin. $N_{ijk}$ follows a Poisson's
distribution with expectation value
$\mu_{ijk} = \left[ b_{jk} + S_{ik} \right] M_j \Delta
t_i \Delta E \epsilon_{jk}$.
The b$_{jk}$ are the background contributions, $M_j$ is the mass of the $j-$th detector,
$\Delta t_i$ is the detector running time during the $i$-th time interval,
$\Delta E$ is the chosen energy bin,
$\epsilon_{jk}$ is the overall efficiency. Moreover, the signal can be written
as $S_{ik} = S_{0,k} + S_{m,k} \cdot \cos\omega(t_i-t_0)$, where $S_{0,k}$ is the constant part of 
the 
signal 
and $S_{m,k}$ is the modulation amplitude.
The usual procedure is to minimize the function $y_k=-2ln({\it\bf L_k}) - const$ for each energy bin;
the free parameters of the fit are the $(b_{jk} + S_{0,k})$ contributions and the $S_{m,k}$
parameter. Hereafter, the index $k$ is omitted when unnecessary.

\begin{figure}[!ht]
\vspace{-0.5cm}
\begin{center}
\includegraphics[width=\textwidth] {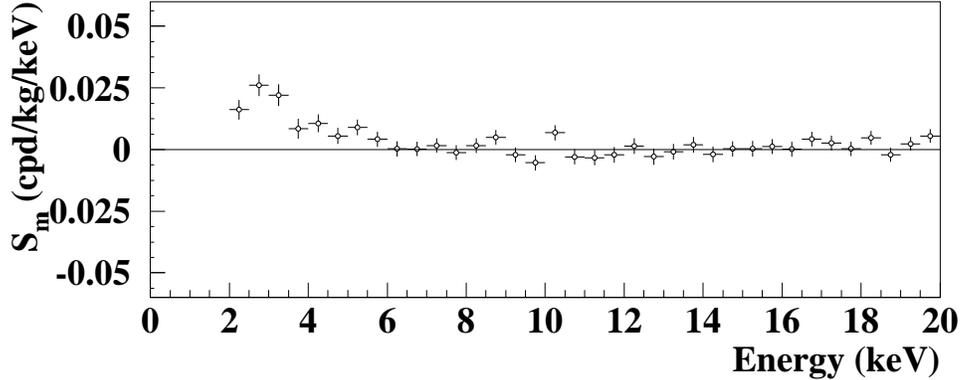}
\end{center}
\vspace{-0.5cm}
\caption{Energy distribution of the $S_{m}$ variable for the
total cumulative exposure 1.17 ton$\times$yr. The energy bin is 0.5 keV.
A clear modulation is present in the lowest energy region,
while $S_{m}$ values compatible with zero are present just above. In fact, the $S_{m}$ values
in the (6--20) keV energy interval have random fluctuations around zero with
$\chi^2$ equal to 27.5 for 28 degrees of freedom.}
\label{sme}
\end{figure}

In Fig. \ref{sme} the obtained $S_{m}$  are shown 
in each considered energy bin (there $\Delta E = 0.5$ keV).
It can be inferred that positive signal is present in the (2--6) keV energy interval, while $S_{m}$
values compatible with zero are present just above. In fact, the $S_{m}$ values
in the (6--20) keV energy interval have random fluctuations around zero with
$\chi^2$ equal to 27.5 for 28 degrees of freedom.
All this confirms the previous analyses.

\vspace{0.5cm}

The method also allows the extraction of the the $S_{m}$ 
values for each detector, for each annual cycle and 
for each energy bin. Thus, following the procedure described in ref. \cite{modlibra},
we have also verified that the S$_{m}$ 
are statistically well distributed in all the six DAMA/LIBRA annual cycles and in all the
sixteen energy bins ($\Delta$E = 0.25 keV in the 2--6 keV energy interval) for each detector.
Moreover, that procedure also allows the definition of a $\chi^2$ 
for each detector; the associated degree of freedom are 16 for the detector restored after the 
upgrade 
in 2008 and 96 for the others. The values of the $\chi^2$/d.o.f. 
range between 0.7 and 1.22 for twenty-four 
detectors, and the observed annual modulation effect is well
distributed in all these detectors at 95\% C.L.. A particular mention is deserved
to the remaining detector whose value is 1.28 exceeding the value corresponding to that C.L.; this 
also is statistically 
consistent, considering that the expected number of detector exceeding this value over twenty-five
is 1.25.
Moreover, the mean value of the 25 $\chi^2/d.o.f.$ is 1.066, slightly larger than expected.
Although this can be still ascribed to statistical fluctuations (see before),
let us ascribe it to a possible systematics. In this case, one would
have an additional error of $\leq 4 \times 10^{-4}$ cpd/kg/keV, if quadratically combined, or
$\leq 5 \times 10^{-5}$ cpd/kg/keV, if linearly combined, to the modulation amplitude
measured in the (2 -- 6) keV energy interval.
This possible additional error: $\leq 4\%$ or $\leq 0.5\%$, respectively, of the
DAMA/LIBRA modulation amplitude
is an upper limit of possible systematic effects.

Among further additional tests, the analysis 
of the
modulation amplitudes as a function of the energy separately for
the nine inner detectors and the remaining external ones has been carried out including the 
DAMA/LIBRA-5,6 data to those already analysed in ref. \cite{modlibra}. 
The obtained values are fully in agreement; in fact,
the hypothesis that the two sets of modulation amplitudes as a function of the
energy belong to same distribution has been verified by $\chi^2$ test, obtaining:
$\chi^2/d.o.f.$ = 3.1/4 and 7.1/8 for the energy intervals (2--4) and (2--6) keV, 
respectively ($\Delta$E = 0.5 keV). This shows that the
effect is also well shared between inner and external detectors. 

Let us, finally, release the assumption of a phase $t_0=152.5$ day in the procedure to 
evaluate the modulation amplitudes from the data of the 1.17 ton $\times$ yr. In this case
alternatively the signal has been written as:

\begin{equation}
S_{ik} = S_{0,k} + S_{m,k} \cos\omega(t_i-t_0) + Z_{m,k} \sin\omega(t_i-t_0) = 
S_{0,k} + Y_{m,k} \cos\omega(t_i-t^*),
\label{eqn} 
\end{equation}

\noindent For signals induced by DM particles one would expect: 
i) $Z_{m,k} \sim 0$ (because of the orthogonality between the cosine and the sine functions); 
ii) $S_{m,k} \simeq Y_{m,k}$; iii) $t^* \simeq t_0=152.5$ day. 
In fact, these conditions hold for most of the dark halo models; however, it is worth noting that 
slight differences can be expected in case of possible contributions
from non-thermalized DM components, such as e.g. the SagDEG stream \cite{epj06} 
and the caustics \cite{caus}.

\begin{figure*}[!ht]
\begin{center}
\vspace{-0.8cm}
\includegraphics[width=0.45\textwidth] {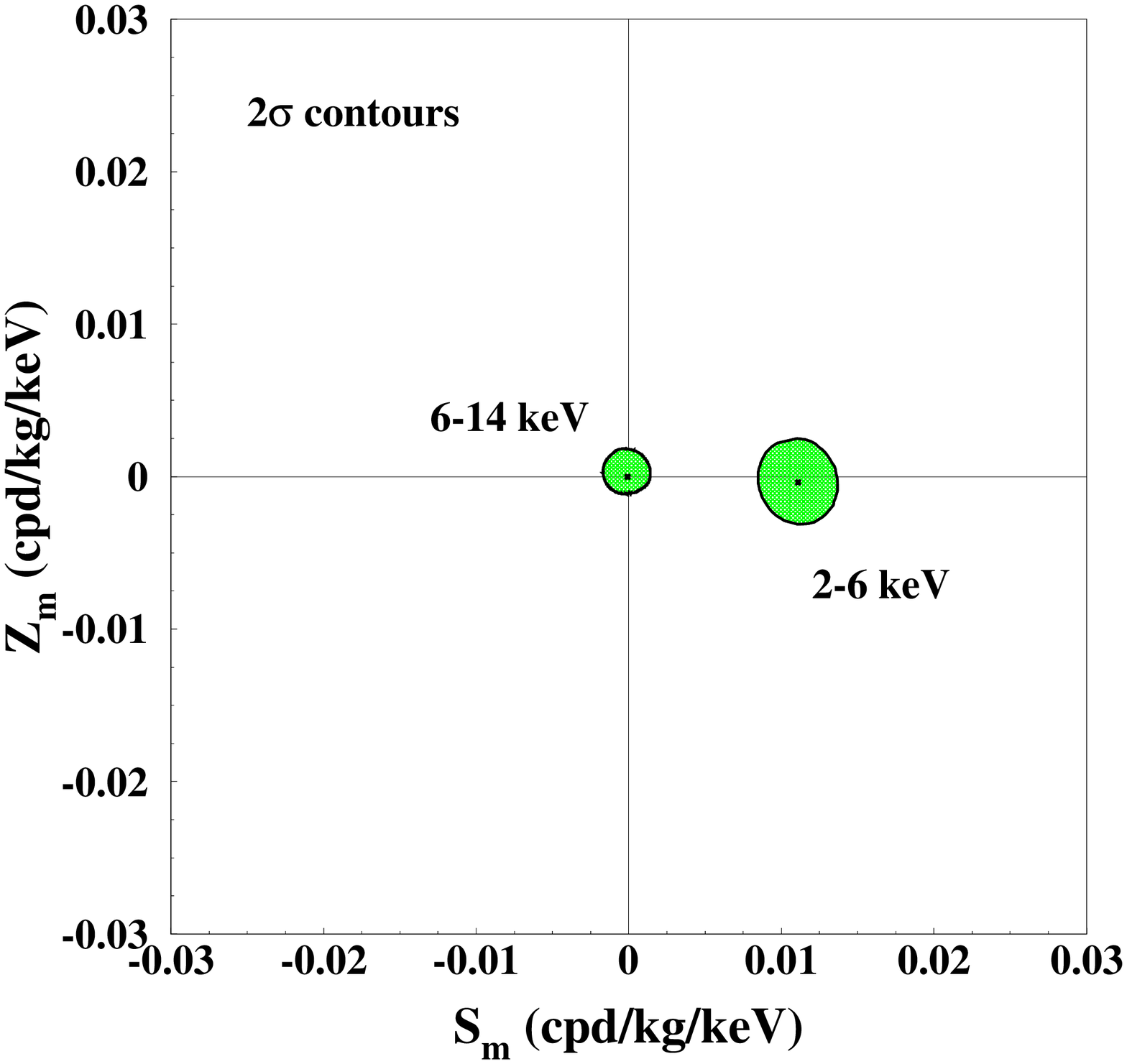}
\includegraphics[width=0.45\textwidth] {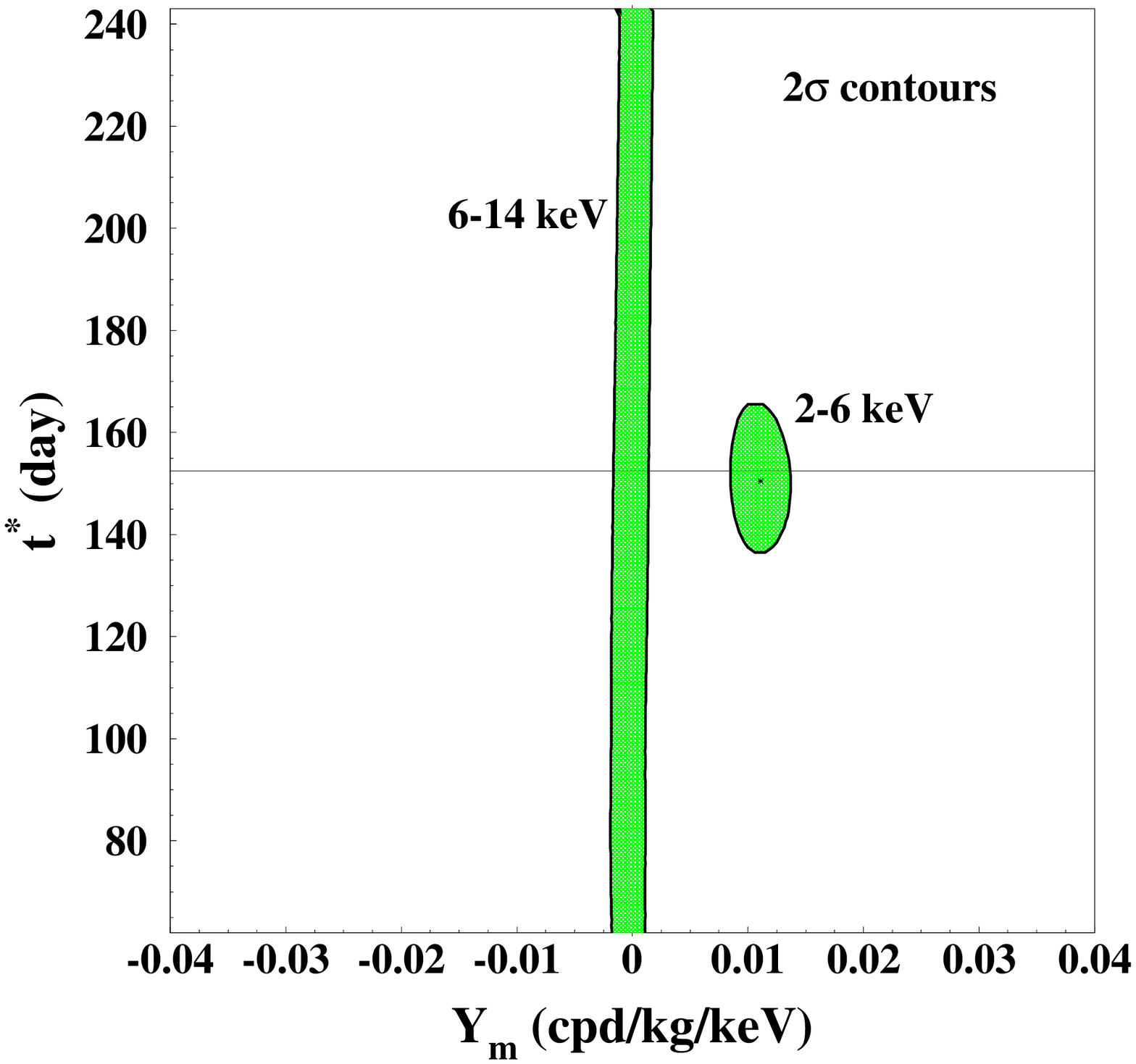}
\end{center}
\vspace{-0.7cm}
\caption{$2\sigma$ contours in the plane $(S_m , Z_m)$ ({\it left})
and in the plane $(Y_m , t^*)$ ({\it right})
for the (2--6) keV and (6--14) keV energy intervals.
The contours have been  
obtained by the maximum likelihood method, considering 
the cumulative exposure of 1.17 ton $\times$ yr.
A modulation amplitude is present in the lower energy intervals 
and the phase agrees with that expected for DM induced signals.}
\label{fg:bid}
\end{figure*}

Fig. \ref{fg:bid}{\it --left} shows the 
$2\sigma$ contours in the plane $(S_m , Z_m)$ 
for the (2--6) keV and (6--14) keV energy intervals and 
Fig. \ref{fg:bid}{\it --right} shows, instead, those in the plane $(Y_m , t^*)$.
Table \ref{tb:bidbf} shows the best fit values for the (2--6) and (6--14) keV energy interval 
($1\sigma$ errors) for  S$_m$ versus  Z$_m$ and  $Y_m$ versus $t^*$.  

\begin{table}[ht]
\vspace{-0.2cm}
\caption{Best fit values for the (2--6) and (6--14) keV energy interval
($1\sigma$ errors) for  S$_m$ versus  Z$_m$ and  $Y_m$ versus $t^*$,
considering the cumulative exposure of 1.17 ton $\times$ yr.
See also Fig. \ref{fg:bid}.}
\vspace{-0.5cm}
\begin{center}
\resizebox{\textwidth}{!}{
\begin{tabular}{|c|c|c|c|c|}
\hline
E     & S$_m$        & Z$_m$        & $Y_m$        & $t^*$  \\
(keV) & (cpd/kg/keV) & (cpd/kg/keV) & (cpd/kg/keV) & (day) \\
\hline
  2--6 &  (0.0111 $\pm$ 0.0013) & -(0.0004 $\pm$ 0.0014) &  (0.0111 $\pm$ 0.0013) & (150.5 $\pm$ 7.0) \\
 6--14 & -(0.0001 $\pm$ 0.0008) &  (0.0002 $\pm$ 0.0005) & -(0.0001 $\pm$ 0.0008) &  undefined       \\
\hline
\hline
\end{tabular}}
\label{tb:bidbf}
\vspace{-0.5cm}
\end{center}
\end{table}

\begin{figure*}[!ht]
\begin{center}
\vspace{-0.5cm}
\includegraphics[width=0.85\textwidth] {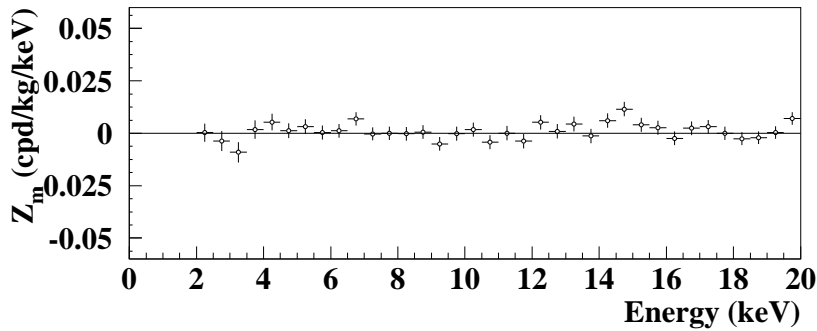}
\end{center}
\vspace{-0.8cm}
\caption{Energy distribution of the $Z_{m}$ variable for the total exposure
(1.17 ton $\times$ yr, DAMA/NaI\&DAMA/LIBRA), once forced 
to zero the contribution of the cosine function in eq. (\ref{eqn}).
The energy bin is 0.5 keV.
The $Z_{m}$ values are expected to be zero in case of
presence of a DM particles' signal with $t^* \simeq t_0 = 152.5$ day.  By the fact, 
the $\chi^2$ test applied to the data supports the hypothesis that the $Z_{m}$ values are simply 
fluctuating around zero; see text.}
\label{fg:zm}
\end{figure*}

Finally, forcing to zero the contribution of the cosine function in eq. (\ref{eqn}),
the $Z_{m}$ values as function of the energy have also been determined
by using the same procedure. The values of $Z_{m}$
as a function of the energy is reported in Fig. \ref{fg:zm}.
Obviously, such values are expected to be zero in case of
presence of a DM signal with $t^* \simeq t_0 = 152.5$ day.  By the fact, 
the $\chi^2$ test 
applied to the data supports the hypothesis that the $Z_{m}$ values are simply 
fluctuating around zero; in fact, for example 
in the (2--14) keV and (2--20) keV energy region the $\chi^2$/d.o.f.
are equal to 21.6/24 and 47.1/36 (probability of 60\% and 10\%), respectively.

\begin{figure*}[!ht]
\begin{center}
\includegraphics[width=0.85\textwidth] {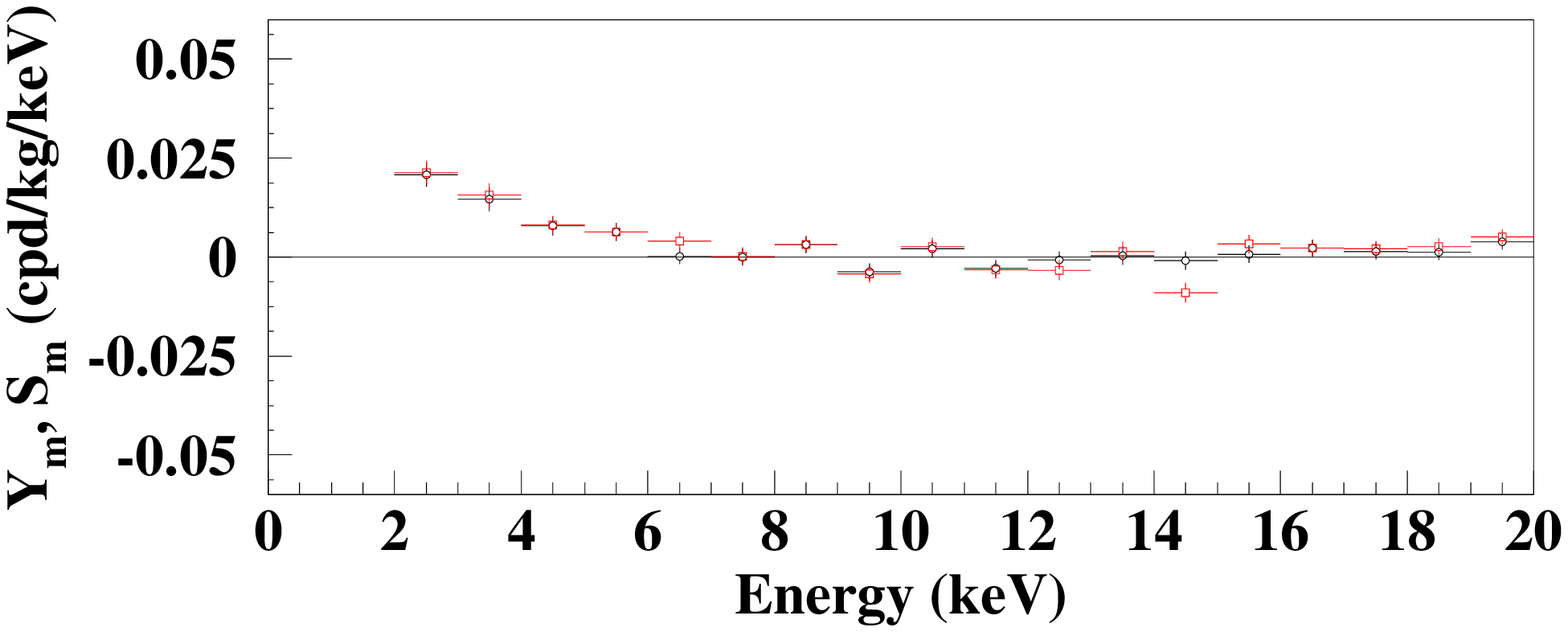}
\includegraphics[width=0.85\textwidth] {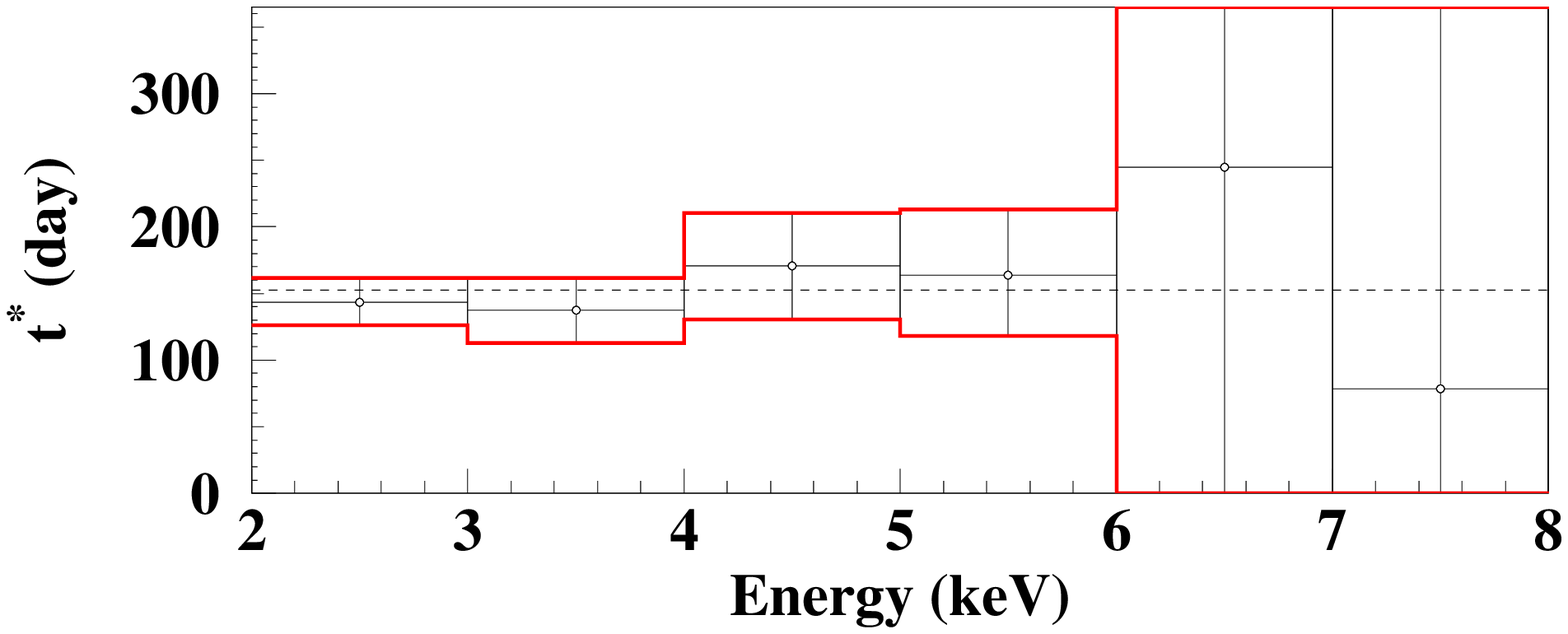}
\end{center}
\vspace{-0.5cm}
\caption{{\em Top:} Energy distributions of the $Y_{m}$ variable (light data points; red colour online)
and of the $S_{m}$ variable (solid data points) for the total exposure
(1.17 ton $\times$ yr, DAMA/NaI\&DAMA/LIBRA). Here, unlike the data of
Fig. \ref{sme}, the energy bin is 1 keV.
{\em Bottom:} Energy distribution of the phase $t^*$ for the total exposure; here 
the errors are at $2\sigma$.
An annual modulation effect is present in the lower energy intervals 
up to 6 keV and the phase agrees with that expected for DM induced signals.
No modulation is present above 6 keV and the phase is undetermined.}
\label{fg:ymts}
\end{figure*}

The behaviours of the $Y_{m}$ and of the phase $t^*$ variables 
as function of energy are shown in Fig. \ref{fg:ymts} 
for the total exposure (1.17 ton $\times$ yr, DAMA/NaI\&DAMA/LIBRA). 
The $Y_{m}$ are superimposed with the $S_{m}$ values 
with 1 keV energy bin (unlike Fig. \ref{sme} where the energy bin is 0.5 keV).
As in the previous analyses, an annual modulation effect is present in the lower energy intervals 
and the phase agrees with that expected for DM induced signals.

These results confirm those achieved by other kinds of analyses.

Sometimes naive statements were put forwards as the fact that
in nature several phenomena may show some kind of periodicity.
It is worth noting that the point is whether they might
mimic the annual modulation signature in DAMA/LIBRA (and former DAMA/NaI), i.e. whether they
might be not only quantitatively able to account for the observed
modulation amplitude but also able to contemporaneously
satisfy all the requirements of the DM annual modulation signature. The same is also for side reactions.
This has already been deeply investigated in ref. \cite{modlibra,perflibra} and references 
therein;
the arguments and the quantitative conclusions, presented there, also 
apply to the DAMA/LIBRA-5,6 data.
Some additional arguments have also been recently addressed in \cite{scineghe09,taupnoz}.

\section{Comments}

The obtained model independent evidence --  at 8.9 $\sigma$ C.L.  over 13 annual cycles --
is compatible with a wide set of scenarios regarding the nature of the DM candidate 
and related astrophysical, nuclear and particle Physics (see 
e.g. 
ref. \cite{RNC,ijmd,ijma,ijma07,chan,wimpele,ldm}, Appendix A of ref. \cite{modlibra} 
and in literature, for example see \cite{Wei01,Botdm,Foot}); 
and many other possibilities are open. Further future works are foreseen.

It is worth recalling that no other experiment exists, whose result can be directly compared in a 
model-independent way with those by DAMA/NaI and DAMA/LIBRA, and that -- more in general --
results obtained with different 
target materials and/or different approaches cannot be 
directly compared among them in a model-independent way.
This is in particular due to the existing experimental and theoretical 
uncertainties,
not last e.g. how many kinds of dark matter particles can exist in the Universe\footnote{
In fact, it is worth noting that, considering the richness in particles of the visible matter which is 
less than 1\%
of the Universe density, one could also expect that the particle part of the Dark Matter in the 
Universe 
may also be multicomponent.},
the nature, the interaction types, the different nuclear and/or atomic correlated aspects,
the unknown right halo model, the 
right DM density, etc. as well as the uncertainties on the values of each one of the 
many involved experimental and 
theoretical parameter/assumption/approximation used in the calculations.
Moreover, some experimental aspects of some techniques used in the 
field have 
also to be addressed \cite{RNC,paperliq,taupnoz}.
Another relevant argument is the methodological robustness \cite{hudson}. 
In particular, the general considerations on comparisons reported in
Appendix A of ref. \cite{modlibra} still hold.
Hence, claims for contradiction have no scientific basis. On the other hand,
whatever possible ``positive'' result has to be interpreted and a large room
of compatibility with DAMA annual modulation evidence is present.

Similar considerations can also be done for the indirect detection searches,
since it does not exist a biunivocal correspondence between the observables in the
direct and indirect experiments. However, if
possible excesses in the positron to electron flux ratio
and in the $\gamma$ rays flux with respect to a modeling of the
background contribution,
which is expected from the considered sources,
might be interpreted -- under some assumptions -- in terms of Dark Matter,
this would also be not in conflict with the effect observed by DAMA experiments.
It is worth noting that different possibilities either considering different background 
modeling or accounting for other kinds of sources can also explain 
the indirect observations \cite{ind}.

Finally, as regards the accelerator searches for new particles beyond
the Standard Model of particle Physics, it is worth noting 
that they can demonstrate the existence of some of the possible DM candidates, but cannot 
credit that a certain particle is the DM solution or the "single" DM solution. 
Moreover, DM candidates and scenarios exist (even e.g. for the neutralino 
candidate) on which accelerators cannot give any information. 
It is also worth noting that for every candidate (including the neutralino) there exist
various different possibilities for the theoretical aspects.
Nevertheless, the results from accelerators will give outstanding and 
crucial complementary information in the field.

\vspace{0.3cm}

A new upgrade of DAMA/LIBRA is foreseen in 2010 with the replacement of all the low 
background PMTs with new ones having higher quantum efficiency;
the main aim is to lower the software energy threshold and, thus, to increase the experimental 
sensitivity and to disentangle -- in the corollary investigation on the candidate particle(s) -- at 
least some of the many possible astrophysical, nuclear and particle Physics scenarios and related 
experimental and theoretical uncertainties.\\

\section{Conclusions}

The new annual cycles DAMA/LIBRA-5,6 have further confirmed a
peculiar annual modulation of the {\it single-hit} events in the (2--6) keV energy region
satisfying the many requests of the DM annual modulation signature; the total 
exposure by former DAMA/NaI and present DAMA/LIBRA is
1.17 ton $\times$ yr. 

In fact, as required by the DM 
annual modulation signature: 1) 
the {\it single-hit} events show a clear cosine-like modulation as expected for the DM signal; 
2) the measured period is equal to $(0.999\pm 0.002)$ yr well compatible with the 1 yr period as 
expected for the DM signal; 3) 
the measured phase $(146\pm7)$ days is well compatible with the roughly $\simeq$ 152.5 days expected for 
the DM signal; 4) the modulation is present only in the low energy (2--6) keV energy 
interval and not in other higher energy regions, consistently with expectation for the DM signal;
5) the modulation is present only in the {\it single-hit} events, while it is absent in the 
{\it multiple-hit} ones as expected for the DM signal;
6) the measured modulation amplitude in NaI(Tl) of the {\it single-hit} events in the (2--6) 
keV energy
interval is: $(0.0116 \pm 0.0013)$ cpd/kg/keV (8.9 $\sigma$ C.L.).
No systematic or side processes able to simultaneously satisfy all the many
peculiarities of the signature and to account for the whole measured modulation
amplitude is available.
Further work is in progress.\\

\end{document}